\documentclass[aps,prl,twocolumn,superscriptaddress]{revtex4-2}

\usepackage{graphicx}
\usepackage{ulem}
\usepackage{color}

\usepackage{xspace}
\usepackage{amsmath}
\usepackage{siunitx}
\usepackage{miller}

\usepackage[version=3]{mhchem}
\usepackage{natbib}
\usepackage{pdfpages}

\makeatletter
\AtBeginDocument{\let\LS@rot\@undefined}
\makeatother

\begin{document}


\title{Rearrangement of orbitals in KAgF$_3$ due to Kugel-Khomskii mechanism: a Neutron diffraction and Density Functional Theory study }

\author{Kacper Koteras}
 \affiliation{%
 University of Warsaw, Center of New Technologies, Żwirki i Wigury 93, 02-089 Warsaw, Poland\\
}%

\author{Sebastian Biesenkamp}
 \affiliation{
 II. Physikalisches Institut, Universität zu Köln, Zülpicher Str. 77, D-50937 Köln, Germany\\
}%

\author{Paolo Barone}
 \affiliation{%
Superconducting and Other Innovative Materials and Devices Institute (SPIN), Consiglio Nazionale delle Ricerche, Area della Ricerca di Tor Vergata, Via del Fosso del Cavaliere 100, I-00133 Rome, Italy\\
}%

\author{Zoran Mazej}
\author{Gašper Tavčar}%
 \affiliation{%
Jožef Stefan Institute, Department of Inorganic Chemistry and Technology, Jamova cesta 39, 1000 Ljubljana, Slovenia\\
}%

\author{Thomas C. Hansen}%
\affiliation{Institut Laue-Langevin, 71 avenue des Martyrs, CS 20156, 38042 Grenoble Cedex 9, France\\}%

\author{José Lorenzana}%
 \email{jose.lorenzana@cnr.it}
 \affiliation{%
Institute for Complex Systems (ISC), Consiglio Nazionale delle Ricerche, Dipartimento di Fisica,
Università di Roma “La Sapienza”, 00185 Rome, Italy\\
}%

\author{Wojciech Grochala}%
 \email{w.grochala@cent.uw.edu.pl}
 \affiliation{%
 University of Warsaw, Center of New Technologies, Żwirki i Wigury 93, 02-089 Warsaw, Poland\\
}%

\author{Markus Braden}%
 \email{braden@ph2.uni-koeln.de}
 \affiliation{
 II. Physikalisches Institut, Universität zu Köln, Zülpicher Str. 77, D-50937 Köln, Germany\\
}%

\date{\today}

\begin{abstract}
The crystal structure of KAgF$_3$ was studied by powder neutron diffraction.
KAgF$_3$ exhibits at all temperatures an orthorhombic symmetry in space group $Pnma$ that allows for several
distortions with respect to the ideal cubic perovskite structure.
At all temperatures there is a strong splitting of Ag-F distances parallel to the $a,c$ planes that documents alternating 
occupation of holes in $x^2-y^2$ and $z^2-y^2$ orbitals.
The orientation of the octahedron elongation and thereby the orbital order flips at a structural phase transition occurring around $T_s=\SI{240}{\kelvin}$ which is accompanied by a suppression of magnetic susceptibility.
The orbital ordering is further enhanced in the low-temperature phase and the twisting of the AgF$_4$ plaquettes forming the antiferromagnetic chains changes.
DFT calculations show an enhancement of the magnetic interaction in the low temperature phase indicating that the transition and the orbital order are partially driven by the Kugel-Khomskii mechanism.
\end{abstract}

\maketitle



The impact of the orbital arrangement on the magnetic interaction is described by the Goodenough-Kanamori-Andersen rules forming the basis of our understanding of magnetic order in transition-metal compounds \cite{khomskii2014}. 
The Kugel-Khomskii model extends this approach by equally treating orbital and magnetic moments so that one may
analyse the interplay of both degrees of freedom \cite{kugel1973}. This model has been first applied to the
magnetic and orbital order of KCuF$_3$ and it seems accepted by now that a large part
of the orbital ordering in this material is induced by the magnetic interactions and the Kugel-Khomskii mechanism \cite{pavarini2008}. In KCuF$_3$, Cu is in its two-valent $3d^9$ configuration with a single hole resulting in a strong Jahn-Teller distortion. 
Electron-phonon coupling combined with an orbitally degenerate state can cause a purely electronically
driven orbital order that is further stabilized by the Kugel-Khomskii mechanism.
In the $a,b$ planes of KCuF$_3$, holes occupy alternatingly $x^2-z^2$ and $y^2-z^2$ orbitals which yields a strong antiferromagnetic interaction along
the $z$ direction and only much weaker exchange parameters perpendicular to it \cite{hutchings1969,hidaka1998,khomskii2014}. The peculiar orbital arrangement thus causes one-dimensional magnetism in this material with a three-dimensional perovskite structure \cite{kugel1973,pavarini2008}. Active orbital degrees of freedom and their impact
on magnetic interaction can be found in numerous compounds such as giant-magnetoresistance manganites \cite{Tokura2006} or multiferroics \cite{Spaldin2019}.

Here we study  KAgF$_3$ \cite{Odenthal1971,mazej2009,kurzydlowski2013}, the $4d$ analogue of KCuF$_3$, which exhibits similar orbital arrangement and similar essentially one-dimensional magnetism mediated by the orbital order \cite{kurzydlowski2017,zhang2011,Wilkinson2023}. In addition, KAgF$_3$ exhibits a structural phase transition around 240\,K, at which the magnetic susceptibility drops indicating an enhancement of the magnetic interaction strength at low temperatures \cite{kurzydlowski2013}. In these combined diffraction and theoretical studies we show that this transition corresponds to an orbital rearrangement at least partially driven by the Kugel-Khomskii effect.


We performed powder neutron diffraction studies on KAgF$_3$ to determine the nuclear structure on the diffractometer D2B at the Institut Laue-Langevin in Grenoble.
A wavelength of \SI{1.593}{\angstrom} was used to obtain high-resolution data sufficiently extending to large $Q$ values. 
We collected data sets at \SI{250}{\kelvin}, \SI{70}{\kelvin}, \SI{40}{\kelvin} and \SI{2}{\kelvin} to cover the structural transition at
$T_s=\SI{240}{\kelvin}$ and the two magnetic transitions deduced from magnetization studies to occur at 35 and 66\,K \cite{kurzydlowski2013}.
Data is available at Ref. \onlinecite{data-5-31-2635} and refinements of structure models were performed with the Fullprof program suite \cite{fullprof}.
We used the same sample as in reference \cite{Wilkinson2023}, where the magnetic structure
of KAgF$_3$ was analyzed.

\begin{figure}
\centering
\includegraphics[width=0.92\columnwidth]{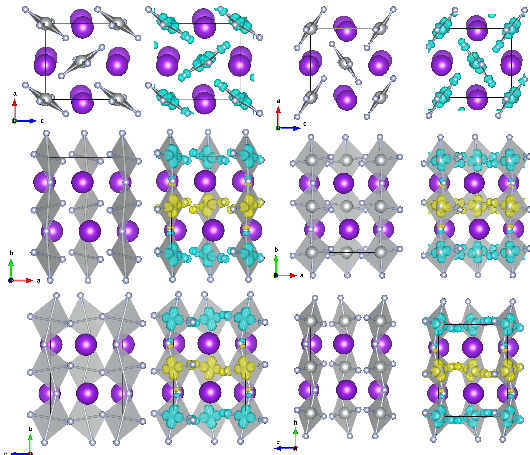}
\includegraphics[width=0.7\columnwidth]{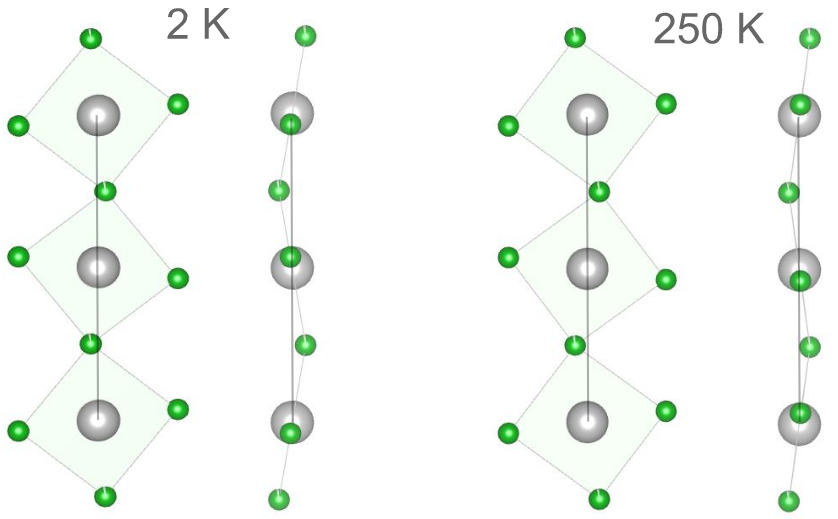}
\caption{ { Crystal structure of KAgF$_3$ as determined with the D2B data at 2\,K (first column) and 250\,K (third column) as well as the two polymorphs obtained in DFT calculations and associated with the low- and high-temperature modifications, respectively (columns 2 and 4). The upper three rows show the projections along the $b$, $c$ and $a$ directions (spacegroup $Pnma$). 
At all temperatures AgF$_6$ octahedra are elongated within the $a,c$ layer so highlighting only the short bonds of the octahedra yields plaquettes (in grey) nearly perpendicular to the layer. 
This elongation alternates within the $a,c$ layer but is inphase along $b$. 
In the columns presenting the DFT calculations, we also show the magnetization densities for up (cyan) and down (yellow) spins highlighting the half-occupied orbital. Most importantly, the plaquettes flip at the structural phase transition. The chains of these AgF$_4$ plaquettes form the magnetic units and are shown in lower row in projections vertical and parallel to the plaquettes.}}
\label{pattern}
\end{figure}

KAgF$_3$, like many distorted perovskite materials, crystallizes in space group $Pnma$
with lattice parameters $a\approx c \approx \sqrt{2}a_{\text{cubic}}$ and $b\approx 2a_{\text{cubic}}$ with respect to the cubic lattice constant of an ideal perovskite. This symmetry allows for various structural distortions, see Fig.~\ref{pattern} ~ \cite{khomskii2014,geller1956,cwik2003,carvajal1998}.
The first column shows the low temperature (LT) structure obtained by neutron diffraction.  
There is a rotation around orthorhombic $b$ (parallel to a Ag-F bond) that is in-phase for octahedra neighboring along $b$ and antiphase in the $a,c$ plane. 
In addition, the octahedra tilt around nearly $a$ with all neighboring octahedra tilting in opposite directions.
Note that this tilt axis is parallel to one of the octahedron edges and that its direction is determined through the space group $Pnma$.
There is only one Ag and one K site but F sites split into F1 bridging two Ag along $b$ and F2 in the $a,c$ plane, see Fig. 1.
Although the ideal perovskite structure is cubic, sometimes a layered character of $a,c$  planes in  $Pnma$ is emphasized \cite{carvajal1998,zhou2005}.
The octahedron in spacegroup $Pnma$ can be distorted, associated with ferro- and antiferroorbital order that determines the magnetic interaction \cite{cwik2003}. 
The antiferroorbital order promoting an $A$-type antiferromagnetic order with ferromagnetic
$a,c$-planes in LaMnO$_3$ is quite famous \cite{carvajal1998} and also KCuF$_3$ exhibits similar orbital order, but note that the space group of KCuF$_3$ is different \cite{hutchings1969,hidaka1998,Lee2012,khomskii2014}. For KCuF$_3$, the Kugel-Khomskii model describing the interplay between orbital and magnetic degrees of freedom was initially developed \cite{kugel1973}. At low temperature KAgF$_3$ exhibits the same antiferroorbital order as LaMnO$_3$ with holes alternatingly occupying
$x^2-y^2$ and $z^2-y^2$ orbitals in the $a,c$ plane (space group $Pnma$ with $b$ the doubled long axis) which results in an $A$-type antiferromagnetic order that
is dominated by the exchange interaction along $b$ \cite{zhang2011,Wilkinson2023}.

\begin{figure}
\includegraphics[width=0.92\columnwidth]{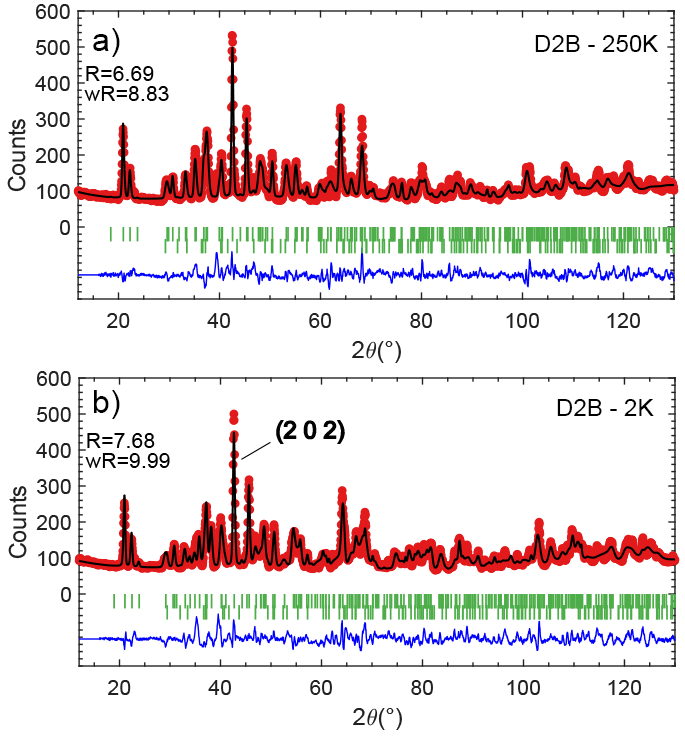}
\caption{ {Rietveld description of the data obtained on the D2B diffractometer at 250\,K (a) and at 2\,K (b). 
Red symbols denote measured intensities, black dots the calculated ones, the blue line the differences, and vertical bars indicate the positions of Bragg reflections of the $Pnma$ KAgF$_3$ and AgF$_2$ phases. }}
\label{fig:rietveld}
\end{figure}

While all previous diffraction studies on KAgF$_3$ found a $Pnma$ structure for the LT phase, there is some 
controversy concerning the high-temperature (HT)
phase above the structural transition \cite{kurzydlowski2013,mazej2009}. 
Neutron diffraction is better suited to analyze the details of the
structural distortion, because the F$^-$  ions more strongly contribute (Fig.~\ref{fig:rietveld}). 
However, with the simple $Pnma$ model, the agreement of structure refinements with the D2B data for KAgF$_3$ is quite poor.
We first focus on the data taken at 2\,K, but the description remained unsatisfactory at all temperatures.
In particular, the very strong (2,0,2) peak exhibits a larger peak height, mostly due to an overestimated peak width. 
Correlated strain corrections \cite{fullprof} improve the description considerably.
In all fits an orthorhombic AgF$_2$ phase \cite{tokar2021} was included, but it represents only a small volume 
fraction (about 6\%), and there are small peaks that remain unexplained (e.g. at \SI{35}{\degree} and \SI{39}{\degree}). 
These peaks could not be indexed with known impurity phases, and they do not strongly depend on temperature,
excluding an origin related to the structural phase transition at $\SI{240}{\kelvin}$.

The fit quality with the data taken at 2\,K improves further when refining anisotropic atomic displacement parameters (ADP), $b_{ij}$, for F2 
(the $a,c$ in-plane F position). The ADPs are strongly enhanced along the [100] direction. Also, an anisotropic Ag ADP improves the fit but less than for F2; here the anomalous enhancement is along [001]. The complete anisotropic treatment at all sites, however, yields several negative values, although it produces the best fit. A model with only two additional parameters is obtained by refining $B_{\text{iso}}$ everywhere and by adding only $b_{11}$ for F2 and $b_{33}$ for Ag, see black curves in Fig. \ref{pattern} and Table 1 in the supplemental material \cite{suppl-mat}.
This is the most simple model that yields a reasonable description of the data taken at 2\,K.

\begin{figure}[t]
\includegraphics[width=0.77\columnwidth]{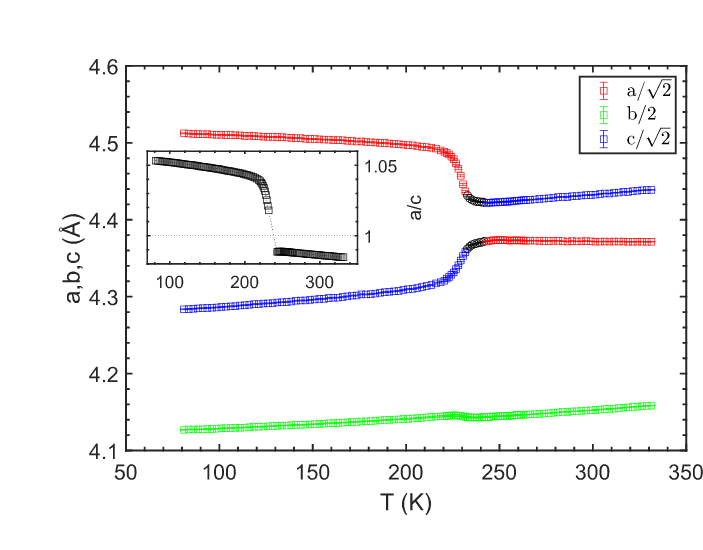}
\caption{ {Temperature dependence of x-ray lattice constants reported in Ref. \cite{kurzydlowski2013} plotted as pseudocubic parameters. The inset shows the $a$/$c$ ratio. Note that the $a$ and $c$ get exchanged at the transition; the black circles in the main plot indicate the fit results near the transition where the crossing of the parameters cannot be properly resolved \cite{note-ac}.  
}}
\label{fig:abcdT}
\end{figure}

\begin{figure}[t]
\includegraphics[width=0.85\columnwidth]{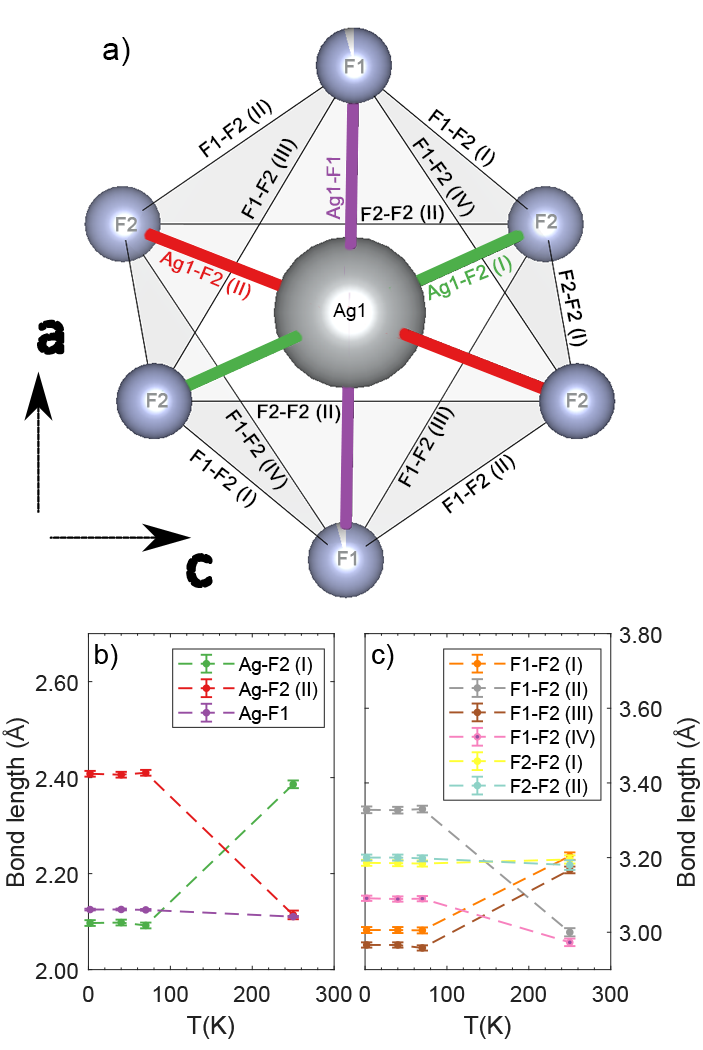}
\caption{ {Sketch of an AgF$_6$ octahedron around the Ag ion at (0.5,0,0.5) illustrating the different bond distances a) and temperature dependence of the
Ag-F b) and F-F distances c) as obtained in the refinements with the $Pnma$ model with two anisotropic ADPs. 
The arrows in (a) indicate the orthorhombic $a$ and $c$ directions. Note that there is a single Ag site, and that the alternating arrangement of long and short bonds within the $a,c$ plane is illustrated in Fig. 1.
Plaquettes of Fig.~\ref{pattern} are defined by the shortest F1-F2 bonds, i.e (I)(III) at low $T$ and (II)(IV) at high $T$. }}
\label{Pnma-dist}
\end{figure}

The refinements at the other studied temperatures below the structural phase transition at 240\,K yield little change and agree with a normal
thermal expansion, however the crystal structure at 250\,K is qualitatively different. 
In Ref.~\cite{kurzydlowski2013} a different symmetry was
proposed for the structure above the structural phase transition: 
In spacegroup $Pcma$ the order of the octahedron rotations is lost along the $b$ direction so that the lattice is not doubled in this direction compared to the ideal perovskite structure. The disappearance of (1,1,1) and (2,1,2) reflections in X-ray diffraction was taken as evidence for this conclusion. 
Also in our neutron data these reflections considerably loose intensity upon heating, however, other rather strong superstructure reflections with odd $k$ index clearly persist, e.g. (2,1,1). Therefore, $Pcma$ can be ruled out. 
The (1,1,1) and (2,1,2) Bragg peaks are not proportional to the rotation angles of the octahedra, and
therefore their suppression only indicates a loss of internal distortion that is however fully compatible with the absence of a symmetry change.

The refinement of the $Pnma$ model with the 250\,K neutron data shows that $c$ is larger than $a$ while the opposite holds at low temperature. 
For the rotation of a regular octahedron one expects the lattice to be elongated along the rotation axis, but the opposite is observed in many three-dimensional and layered perovskites \cite{okeefe1977,braden1994,cwik2003,zhou2005}. 
In $Pnma$ the tilting occurs around an axis parallel to the octahedron edge, which, due to the $b$-axis rotation, is shifted by this angle of $\sim$\SI{10.5}{\degree} away from $a$.
With the proper assessment of $a$ and $c$ in space group $Pnma$ \cite{note-ac} we obtain the temperature dependence of the lattice constants determined by x-ray diffraction shown in Fig.~\ref{fig:abcdT}.

Most importantly also the elongation of the octahedra along an in-plane bond flips at this phase transition.
For an ideal crystal structure without any octahedron rotation around $b$, this flipping just signifies a phase shift, but with the strong $b$-axis rotation of the octahedra
the elongation flip signifies a distinct arrangement, which, however, does not imply a different space group. 
At low temperature, the elongations at all Ag places are rotated towards the $a$ direction (forming an angle of \SI{34}{\degree} with $a$) while they are
rotated towards $c$ at 250\,K (angle of \SI{36}{\degree}). 
The flop of the octahedron elongation signifies a change of the half-occupied $e_g$ orbitals ($x^2-z^2$ versus $y^2-z^2$) that lay in the plaquettes formed by the four short bonds.
These distinct schemes  are illustrated in Fig.~\ref{pattern} and indicate that the orbital order changes at $T_s$. While the antiferroorbital order in the $a,c$ planes as well as the ferroorbital stacking along $b$ do not change between the LT and HT phases, it is the embedding of the orbital order in the rotation and tilting distortions that gets modified and leads to a stronger local octahedron distortion. 
This subtle change, however, possesses a strong impact on the magnetic interaction, as will be shown below. 

\begin{figure}
\includegraphics[width=0.85 \columnwidth]{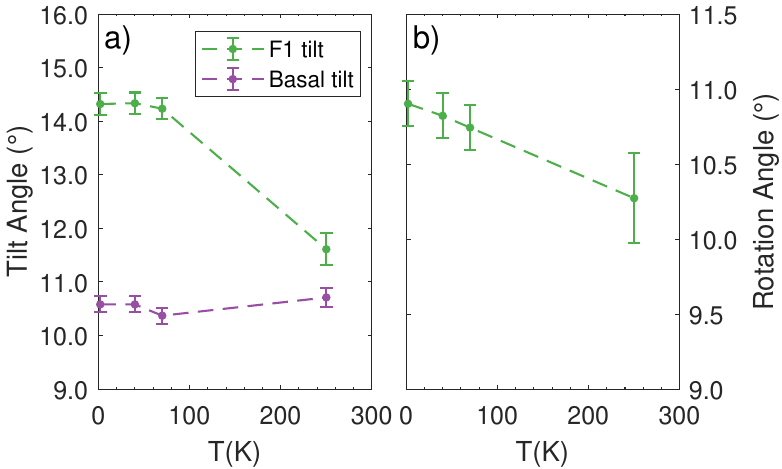}
\caption{ {Temperature dependence of the tilt (a) and rotation (b) angles, as obtained in the refinements with the $Pnma$ model with two anisotropic ADPs. }}
\label{tilt-rotation}
\end{figure}

The structural phase transition in KAgF$_3$ can thus be identified with a change of orbital order. The first-order character of this phase transition
is in agreement with both phases showing $Pnma$ symmetry, which enforces the pattern of the elongation order that is alternating in the $a,c$ layer and 
inphase along $b$. However, the $Pnma$ symmetry does not fix the direction of the elongation and thus also the direction of the orbital order can vary.

Figure~\ref{Pnma-dist} presents the Ag-F and F-F distances obtained in the refinements with the $Pnma$ model
with two anisotropic ADPs and a sketch of the AgF$_6$ octahedron. 
The Ag lies at an inversion symmetry center so that there are only three distinct Ag-F distances and 6 distinct F-F octahedron edges. 
Note that the arrangement of short and long Ag-F2 bonds alternates for neighboring Ag sites in an $a,c$ layer 
yielding antiferro-orbital inplane order.  
More importantly, Fig. \ref{Pnma-dist} shows that, at the structural transition, the long and short Ag-F2 distances are inverted. In consequence, also the long and short F1-F2 distances become inverted. Since the shorter F1-F2 bonds form the plaquettes containing the half-filled orbital, see Fig.~\ref{pattern}, the orbital order qualitatively changes. The splitting of these distances is furthermore enhanced at low temperatures (by about 10\% \ for Ag-F2) signifying enhanced orbital order.
On the other hand, there is only a tiny splitting in the two in-plane F2-F2 octahedron edges as opposed to the large splitting seen in systems with ferroorbital ordering \cite{cwik2003}.  This tiny edge splitting also gets inverted at the structural transition.

The tilt and rotation angles of the octahedra exhibit only a small temperature dependence, see Fig.~\ref{tilt-rotation}.
Note that the tilt of the octahedron around the $a$ axis can be determined with the apical F1 site and with the basal plane.
The difference in these values indicates some further distortion of the octahedra that can be ascribed to the direct F1 K interaction.
The rotation of the octahedron around $b$ is deduced from the in-plane F2 site only. Although these temperature-dependent changes
may be relevant for the orbital rearrangement they do not seem to reflect the main cause. 

For KCuF$_3$ it was reported that the
Kugel-Khomskii coupling considerably contributes to the orbital ordering \cite{pavarini2008,Lee2012}. Since the impact of magnetic superexchange increases upon
cooling, it can explain the temperature-driven rearrangement of orbital order in KAgF$_3$. This picture also agrees with the enhanced elongation observed at low temperatures, where the impact of magnetic exchange increases.

Figure~\ref{pattern} and Table I in the supplemental material \cite{suppl-mat} show the temperature dependence of the extra anisotropic ADPs obtained with the $Pnma$ model. The values
are way too large (about an order of magnitude) to agree with usual harmonic lattice dynamics \cite{braden2001} but clearly point to disorder or strong anharmonicity.
At low temperatures, the values indicate root-mean-square displacements of \SI{0.18}{\angstrom} and \SI{0.22}{\angstrom} for Ag and F2, respectively.
The disorder at the F2 site further increases to a root-mean-square value of \SI{0.26}{\angstrom} at \SI{250}{\kelvin}, where it points along the $c$ direction. 

The disorder seems to arise from local defects, because it follows the orbital rearrangement and 
because it only quantitatively changes at $T_s$ besides the inversion of the directions of extra ADPs. 
Stacking faults or a competing scheme of stacking order along $b$ appear most likely, because the strong variation of long and
short distances within the $a,c$ layers imposes long-range order.
The in-plane ordering and the disorder resemble the orbital order and the observation of two distinct orbital arrangements in KCuF$_3$ \cite{hidaka1998}. In its majority phase, KCuF$_3$ 
crystallizes in the tetragonal spacegroup $I4/mcm$, which allows for an octahedron rotation around $c$ and for the same inplane
bond-distance splitting or orbital pattern that we find in the $a,c$ layers of  KAgF$_3$ (note that due to the distinct symmetries distinct planes are considered, $a,b$ planes in KCuF$_3$ and $a,c$ planes in KAgF$_3$). But the main orbital arrangement along the
vertical axis is perpendicular (or alternating) for KCuF$_3$, while we observe a parallel alignment in KAgF$_3$, see Fig.~\ref{pattern}. However, KCuF$_3$ can also show
a minority orbital and bond distance pattern that corresponds to the ordering scheme in KAgF$_3$.
The distinct patterns can be clearly attributed to the distinct stacking sequences of the octahedron rotations in
spacegroups $I4/mcm$ and $Pnma$, which are antiphase in the former and inphase in the latter. 
The disorder evidenced by the extra F2 ADP can be attributed to a similar stacking disorder of orbital order, which
is further supported by its analysis with an extra F2 site, see supplemental information \cite{suppl-mat}.

\begin{table}
\caption{\label{tab:table_DFT_vs_exp} Comparison of lattice parameters (in \AA) for the LT and HT phases coming from neutron diffraction and from DFT calculations. For the experiment, we report the values at 2K and 250K as representative of the LT and HT phases respectively.  The last two columns compare the calculated magnetic interaction parameters (in meV) for the two phases and with an experimental estimation from Ref.~\cite{kurzydlowski2013}.
}
\begin{ruledtabular}
\begin{tabular}{m{0.9cm}cccccccc}
 &&a  &b  &c &$J_{1D}$ &$J_{perp}$ \\ \hline
HT	
      & exp. & 6.1864(10)  & 8.2683(7)  & 6.2413(8) & -97& - \\
      & theor. & 6.139  & 8.169  & 6.159 & -106.6 & +1.69 \\
\\
LT      & exp. & 6.3611(10)  & 8.2362(7)  & 6.0677(6) & ND & - \\
      & theor. & 6.309  & 8.180  & 6.018 & -114.1 & +1.54 \\
\end{tabular}
\end{ruledtabular}
\end{table}


In order to understand the structural transition, we have performed DFT computations, for technical details see the supplemental information \cite{suppl-mat}. 

Remarkably, we have found that two possible structures minimize the DFT energy with \textit{Pnma} 
symmetry which correspond closely to the experimental LT and the HT structure as shown in Fig.~\ref{pattern} (and Table I in the supplemental material \cite{suppl-mat}). The structure of the two phases can be rationalized as driven by the collective Jahn-Teller distortions. Both phases show no imaginary phonon modes (cf. Supplementary Information \cite{suppl-mat}) so they correspond to genuine local minima \cite{togo2023}. 

The fact that the HT phase is found in DFT is surprising because, usually, above a transition temperature entropic effects determine the structure which does not necessarily correspond to a minimum of the $T=0$ DFT functional. Here, entropic effects tip the balance between two pre-existent minima  \cite{suppl-mat}.

The LT phase is characterized by $a > c$, while for the HT phase $a < c$. This is consistent with the different ordering of long axes of the AgF$_6$ octahedra in both phases, as seen in neutron diffraction. Simultaneously, $b$ is slightly larger for the LT phase than for the HT phase (formally at 0 K); note that this anomaly at the phase transition ($T_s=240$ K) is clear 
in x-ray diffraction (Ref.~\cite{kurzydlowski2013} and Fig.~\ref{fig:abcdT}). 
Similar results were obtained with neutrons (not shown).


Magnetic superexchange constants were calculated using the broken-symmetry method mapping calculated energy values of different magnetic configurations in DFT to 
a Heisenberg Hamiltonian: 
$ H= -\sum_{\langle i,j\rangle } J_{ij} {\bf S}_i. {\bf S}_j $, where the coupling constants  $J_{ij}$ correspond to bonds. 
By calculating the total energy of different magnetic configurations with spins aligned along $z$, it is possible to derive the superexchange coupling constants. Three states were used for each structure in order to derive the superexchange constant along [AgF$^+$] chains ($J_{1D}$) and diagonally in [AgF$_2$] planes ($J_{perp}$).

Analysis of the magnetic superexchange interactions in both polymorphs reveals another important difference (Table I). The interchain superexchange is similar and weakly ferromagnetic for both forms (ca. 1.5-1.7 meV), which agrees with
the A-type magnetic structure with ferromagnetic $a,c$ layers \cite{Wilkinson2023}.
In contrast, the absolute value of the antiferromagnetic intrachain $J_{1D}$ is as much as 7 percent (or 7.5 meV) larger for the LT form. This is a huge contribution, which surpasses the difference of electronic energies of both forms at 0 K by the factor of five. This result reconfirms the key role of magnetic superexchange for relative stability of both forms, and thus the Kugel-Khomskii mechanism for the LT/HT phase transition.

Several structural aspects may contribute to the increase of intrachain interaction in the LT phase. The orbital ordering is enhanced due to the larger splitting of the Ag-F2 distances, and the twisting of the plaquettes in the magnetic chains is modified as it is illustrated in Fig. 1. In KCuF$_3$ \cite{pavarini2008} and in other materials with coupled orbital and magnetic order \cite{pavarini2010} it took a long time to quantitatively access the impact of the Kugel-Khomskii mechanism in the orbital order, that otherwise is caused by electron-phonon coupling acting on orbital degeneracy.
In contrast for KAgF$_3$, the orbital rearrangement associated with a drop in susceptibility \cite{kurzydlowski2013} and with an increase of the magnetic interaction unambigously documents its impact.


In conclusion, we have studied the nuclear structure of KAgF$_3$ as a function of temperature using powder neutron diffraction, x-rays and DFT based computations. KAgF$_3$ can be seen as a system with two nearly degenerate polymorphs which compete as function of temperature. Both polymorphs are stable minima of the DFT functional, and subtle effects can tip the balance between the two states.  
The structural phase transition at \SI{240}{\kelvin} is identified as a rearrangement of orbital order and of the associated octahedron elongation.
The analysis of both structural polymorphs by DFT calculations indicates that the structural
arrangement in the low-temperature phase exhibits a larger magnetic exchange interaction, which agrees with the drop of the magnetic susceptibility at this transition. Lowering the temperature, the transition can be seen as driven by a loss of lattice entropy compensated by a gain in magnetic energy of the LT phase. Such gain in magnetic energy can be traced back to the orbital rearrangement, as expected from a  Kugel-Khomskii mechanism.  Therefore,  KAgF$_3$ is a unique system where magnetism  determines the structural stability.


\begin{acknowledgments}

Research was carried out with the use of CePT infrastructure financed by the European Union—the European Regional Development Fund within the Operational Programme “Innovative economy” for 2007–2013 (Grant No. POIG.02.02.00-14-024/08-00). The Polish authors are grateful to Narodowe Centrum Nauki (Poland) for support (Maestro, Grant No. 2017/26/A/ST5/00570). The Italian authors acknowledge financial support from the Italian MIUR through Projects No. PRIN 2017Z8TS5B and No. PRIN 20207ZXT4Z. The Slovenian authors acknowledge the financial support of the Slovenian Research and Innovation Agency (Research Core Funding No. P1-0045; Inorganic Chemistry and Technology). The German authors acknowledge the Deutsche Forschungsgemeinschaft (DFG, German Research Foundation) - Project number 277146847 - CRC 1238, project B04. W.G. is grateful to the Interdisciplinary Center for Mathematical and Computational Modelling, University of Warsaw, for the availability of high performance computing resources (Okeanos, Topola) within Project No. GA83-34. 
  J.L. acknowledges hospitality from  Kavli Institute for Theoretical Physics, Santa Barbara. This research was supported in part by the National Science Foundation under Grants No. NSF PHY-1748958 and PHY-2309135.
The authors thank D. Khomskii for stimulating discussions.
KK and SB contributed equally to this work.
\end{acknowledgments}

\nocite{apsrev41Control}
\bibliographystyle{apsrev4-2}

%


\begin{thebibliography}{28}
\bibitem [{not()}]{note-ac}%
	\BibitemOpen
	\href@noop {} {}\bibinfo {note} {We describe the structure in space-group
		$Pnma$ at all temperatures, which signifies that the tilt axis of the AgF$_6$
		octahedron is kept nearly parallel to the $a$ direction. If one prefers to
		keep $a>c$ in the HT phase, the space-group setting must be changed to $Pcmn$
		illustrating the first-order character of the phase transition. The
		low-temperature orthorhombic splitting with $a$$>$$c$ in $Pnma$ points to a
		more rigid rotation of the octahedron basal plane.}\BibitemShut {Stop}%
\end{thebibliography}

\clearpage
 \hspace*{-2cm} 
 \vspace*{-1cm}
\includepdf[pages=1]{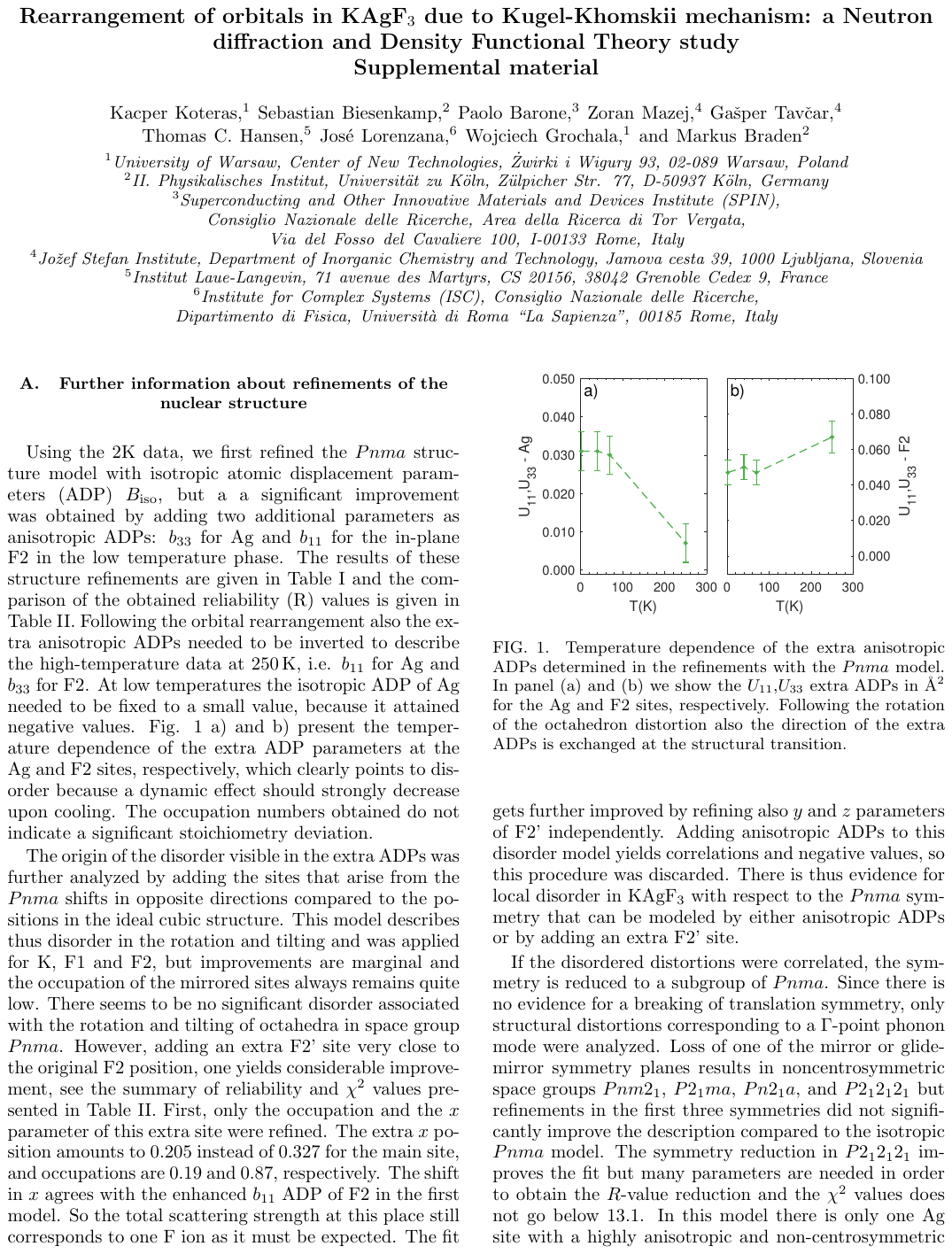} 

\clearpage
\hspace*{-2cm} 
\vspace*{-1cm}
\includepdf[pages=2]{nucl-struc-KAgF3-SMv1.pdf} 

\clearpage
\hspace*{-2cm} 
\vspace*{-1cm}
\includepdf[pages=3]{nucl-struc-KAgF3-SMv1.pdf} 

\clearpage
\hspace*{-2cm} 
\vspace*{-1cm}
\includepdf[pages=4]{nucl-struc-KAgF3-SMv1.pdf} 

\clearpage
\hspace*{-2cm} 
\vspace*{-1cm}


\section*{Supplementary material}

# Rearrangement of orbitals in KAgF$_3$ due to Kugel-Khomskii mechanism: a Neutron diffraction and Density Functional Theory study
## Supplemental material

Kacper Koteras,[1] Sebastian Biesenkamp,[2] Paolo Barone,[3] Zoran Mazej,[4] Gašper Tavčar,[4] Thomas C. Hansen,[5] José Lorenzana,[6] Wojciech Grochala,[1] and Markus Braden[2]

[1] *University of Warsaw, Center of New Technologies, Żwirki i Wigury 93, 02-089 Warsaw, Poland*
[2] *II. Physikalisches Institut, Universität zu Köln, Zülpicher Str. 77, D-50937 Köln, Germany*
[3] *Superconducting and Other Innovative Materials and Devices Institute (SPIN), Consiglio Nazionale delle Ricerche, Area della Ricerca di Tor Vergata, Via del Fosso del Cavaliere 100, I-00133 Rome, Italy*
[4] *Jožef Stefan Institute, Department of Inorganic Chemistry and Technology, Jamova cesta 39, 1000 Ljubljana, Slovenia*
[5] *Institut Laue-Langevin, 71 avenue des Martyrs, CS 20156, 38042 Grenoble Cedex 9, France*
[6] *Institute for Complex Systems (ISC), Consiglio Nazionale delle Ricerche, Dipartimento di Fisica, Università di Roma "La Sapienza", 00185 Rome, Italy*

## A. Further information about refinements of the nuclear structure

Using the 2K data, we first refined the *Pnma* structure model with isotropic atomic displacement parameters (ADP) $B_{iso}$, but a a significant improvement was obtained by adding two additional parameters as anisotropic ADPs: $b_{33}$ for Ag and $b_{11}$ for the in-plane F2 in the low temperature phase. The results of these structure refinements are given in Table I and the comparison of the obtained reliability (R) values is given in Table II. Following the orbital rearrangement also the extra anisotropic ADPs needed to be inverted to describe the high-temperature data at 250 K, i.e. $b_{11}$ for Ag and $b_{33}$ for F2. At low temperatures the isotropic ADP of Ag needed to be fixed to a small value, because it attained negative values. Fig. 1 a) and b) present the temperature dependence of the extra ADP parameters at the Ag and F2 sites, respectively, which clearly points to disorder because a dynamic effect should strongly decrease upon cooling. The occupation numbers obtained do not indicate a significant stoichiometry deviation.

The origin of the disorder visible in the extra ADPs was further analyzed by adding the sites that arise from the *Pnma* shifts in opposite directions compared to the positions in the ideal cubic structure. This model describes thus disorder in the rotation and tilting and was applied for K, F1 and F2, but improvements are marginal and the occupation of the mirrored sites always remains quite low. There seems to be no significant disorder associated with the rotation and tilting of octahedra in space group *Pnma*. However, adding an extra F2' site very close to the original F2 position, one yields considerable improvement, see the summary of reliability and $\chi^2$ values presented in Table II. First, only the occupation and the $x$ parameter of this extra site were refined. The extra $x$ position amounts to 0.205 instead of 0.327 for the main site, and occupations are 0.19 and 0.87, respectively. The shift in $x$ agrees with the enhanced $b_{11}$ ADP of F2 in the first model. So the total scattering strength at this place still corresponds to one F ion as it must be expected. The fit

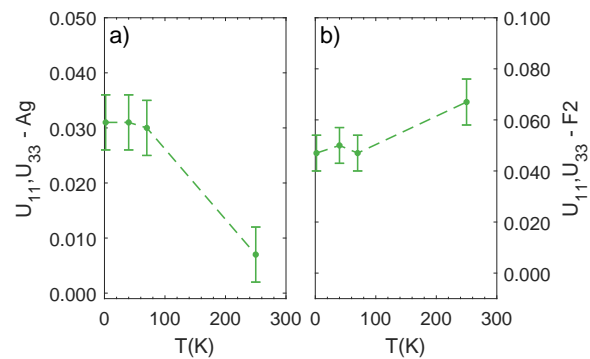

FIG. 1. Temperature dependence of the extra anisotropic ADPs determined in the refinements with the *Pnma* model. In panel (a) and (b) we show the $U_{11}$,$U_{33}$ extra ADPs in Å$^2$ for the Ag and F2 sites, respectively. Following the rotation of the octahedron distortion also the direction of the extra ADPs is exchanged at the structural transition.

gets further improved by refining also $y$ and $z$ parameters of F2' independently. Adding anisotropic ADPs to this disorder model yields correlations and negative values, so this procedure was discarded. There is thus evidence for local disorder in KAgF$_3$ with respect to the *Pnma* symmetry that can be modeled by either anisotropic ADPs or by adding an extra F2' site.

If the disordered distortions were correlated, the symmetry is reduced to a subgroup of *Pnma*. Since there is no evidence for a breaking of translation symmetry, only structural distortions corresponding to a Γ-point phonon mode were analyzed. Loss of one of the mirror or glide-mirror symmetry planes results in noncentrosymmetric space groups $Pnm2_1$, $P2_1ma$, $Pn2_1a$, and $P2_12_12_1$ but refinements in the first three symmetries did not significantly improve the description compared to the isotropic *Pnma* model. The symmetry reduction in $P2_12_12_1$ improves the fit but many parameters are needed in order to obtain the R-value reduction and the $\chi^2$ values does not go below 13.1. In this model there is only one Ag site with a highly anisotropic and non-centrosymmetric

TABLE I. Results of the Rietveld refinements with the anisotropic $Pnma$ model, basing on the data collected on D2B. For all temperatures, an impurity phase of $AgF_2$ was included in the refinement. ADPs are transformed to $U_{iso}$ and $U_{ii}$ because of the simpler interpretation as mean square displacements and the values are given in Å$^2$.

|  |  | x | y | z | Occ | $U_{iso}$ | $U_{11}$ | $U_{33}$ |
|---|---|---|---|---|---|---|---|---|
| 250 K | Ag | 0 | 0 | 0 | 1 | 0.0022(15) | 0.007(5) | - |
|  | K1 | 0.038(2) | 0.25 | 0.479(2) | 1.10(2) | 0.031(4) | - | - |
|  | F1 | 0.485(2) | 0.25 | 0.5664(10) | 1.052(19) | 0.030(3) | - | - |
|  | F2 | 0.2780(15) | 0.4641(6) | 0.1910(12) | 1.006(16) | 0.022(3) | - | 0.067(9) |
|  | $a$=6.1864(10), $b$=8.2683(7), $c$=6.2413(8), R=6.69, wR=8.83 | | | | | | | |
| 70 K | Ag | 0 | 0 | 0 | 1 | 0.0100(15) | - | 0.030(5) |
|  | K1 | 0.0475(17) | 0.25 | 0.4881(18) | 1.00(2) | 0.012(3) | - | - |
|  | F1 | 0.4760(13) | 0.25 | 0.5823(12) | 0.967(18) | 0.002(2) | - | - |
|  | F2 | 0.3110(9) | 0.4652(5) | 0.2219(10) | 1.004(18) | 0.016(2) | 0.047(7) | - |
|  | $a$=6.3587(11), $b$=8.2367(8), $c$=6.0695(6), R=7.30, wR=9.58 | | | | | | | |
| 40 K | Ag | 0 | 0 | 0 | 1 | 0.0102(16) | - | 0.031(5) |
|  | K1 | 0.0477(17) | 0.25 | 0.4884(18) | 1.01(2) | 0.014(3) | - | - |
|  | F1 | 0.4754(13) | 0.25 | 0.5828(12) | 0.970(19) | 0.002(2) | - | - |
|  | F2 | 0.3106(9) | 0.4645(5) | 0.2211(10) | 1.001(18) | 0.017(2) | 0.050(7) | - |
|  | $a$=6.3612(11), $b$=8.2364(7), $c$=6.0681(6), R=7.58, wR=9.92 | | | | | | | |
| 2 K | Ag | 0 | 0 | 0 | 1 | 0.0105(16) | - | 0.031(5) |
|  | K1 | 0.0486(17) | 0.25 | 0.4886(18) | 1.01(2) | 0.011(3) | - | - |
|  | F1 | 0.4754(13) | 0.25 | 0.5827(12) | 0.967(18) | 0.002(2) | - | - |
|  | F2 | 0.3111(9) | 0.4645(5) | 0.2208(10) | 0.994(18) | 0.016(2) | 0.047(7) | - |
|  | $a$=6.3611(10), $b$=8.2362(7), $c$=6.0677(6), R=7.68, wR=9.99 | | | | | | | |

TABLE II. Comparison of different models to describe the high resolution D2B data at low temperature. We refined either isotropic or anisotropic ADPs and occupation numbers of K and F sites.

| refinement model | $R_p$ | $R_{wp}$ | $\chi^2$ |
|---|---|---|---|
| $Pnma$ with iso. ADP and occ. free | 7.92 | 10.2 | 13.8 |
| $Pnma$ all iso. $b_{11}$-F2 $b_{33}$-Ag [a] | 7.65 | 9.96 | 13.1 |
| $Pnma$ all aniso. | 7.61 | 9.92 | 13.0 |
| $Pnma$ all iso. $b_{11}$-F2 $b_{33}$-Ag [b] | 7.68 | 9.99 | 13.2 |
| $Pnma$ extra F2 site with $x$ free | 7.69 | 9.99 | 13.2 |
| $Pnma$ extra F2 site with $x, y, z$ free | 7.61 | 9.93 | 13.0 |
| $P112_1/a$ only $x$-F2' free | 7.69 | 9.99 | 13.2 |
| $P112_1/a$ only $x, y, z$-F2' free | 7.61 | 9.93 | 13.0 |
| $P112_1/a$ $x, y, z$ occ.-F2' free | 7.66 | 9.87 | 12.9 |
| $P112_1/a$ $x, y, z$ $U_{iso}$ occ.-F2' free | 7.61 | 9.78 | 12.7 |
| $P2_12_12_1$ $xyz$ occ.-F2' $x, y, z$-Ag $y$-K $y$-F1 | 7.61 | 9.94 | 13.1 |

[a] $B_{iso}$ Ag goes negative
[b] $B_{iso}$ Ag fixed to .01

surrounding. For example the two long in-plane bond distances amount to 2.63 Å and 2.23 Å. In view of the many additional parameters and the resulting strong deformations, this model seems little justified. The Γ-point distortions yielding a centrosymmetric symmetry correspond to space groups $P2_1/n11$, $P112_1/a$, and $P12_1/m1$, where we use nonstandard settings in order to keep the original lattice (the 1 indicates the identity operation to identify the orientation of the monoclinic axis). The space groups are monoclinic with the monoclinic axis being $a$, $c$ or $b$, respectively. Since the (2,0,2) reflection seems to be sharper than average, $P12_1/m1$ can be discarded because the monoclinic angle would yield a splitting of this peak. The other two groups were tested. Only the refinement of the model with space group $P112_1/a$ yields a significant improvement of the fit compared to the isotropic $Pnma$ model. In this symmetry there are two Ag sites that occupy alternatingly the planes perpendicular to $b$ and each site has only F2 or F2' neighbors. This symmetry thus requires some stacking order of electronic and orbital degrees of freedom. Since also this model does not yield a clear improvement compared to the disorder models it seems to represent just another - the third - possibility to handle the disorder. We will discuss it further below. It cannot be fully excluded that the true order of this distortion is overlooked in the analysis and that there is breaking of translation symmetry, which could explain the low occupation of the extra site in the refinement. However, disorder induced by some defects seems to be more likely as the disorder does not disappear with increasing temperature, see Fig. 1.

Table II compares the reliability factors of the refinements with different structural models. The model in space group $Pnma$ where ADPs are treated in an

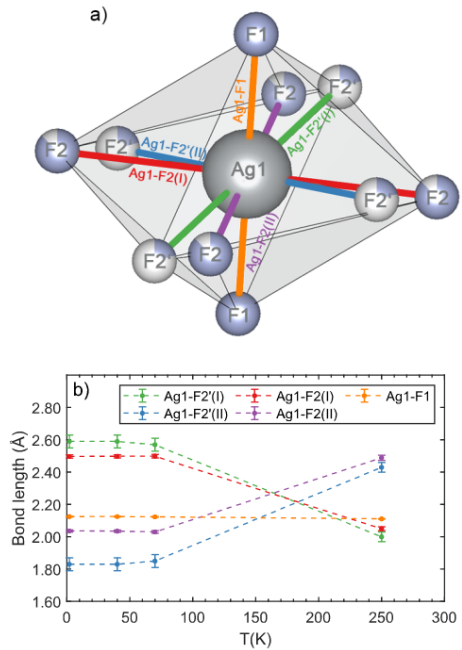

FIG. 2. Sketch of the octahedron with the two F2 and F2' sites and temperature dependence of the bond distances. The blue shading of the F2 sites indicates their occupation numbers, which add to nearly one.

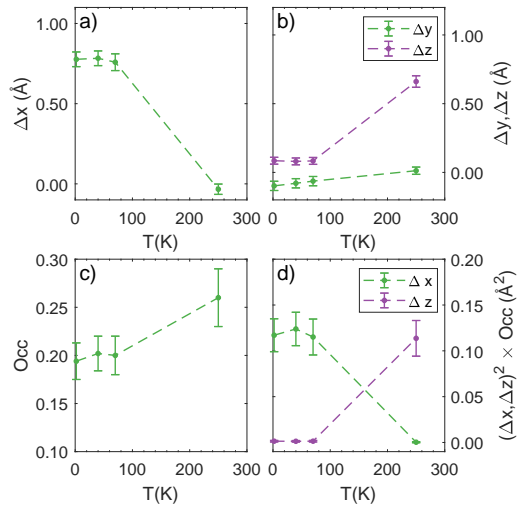

FIG. 3. Temperature dependence of the parameters describing the disorder model with an extra F2' site. Panel (a) shows the difference in the $x$ parameter and panel (b) those for $y$ and $z$ multiplied by the lattice constant to yield absolute units. Panel (c) presents the occupation of the extra site which increases with temperature. Panel (d) shows the product of the squares of the differences and the occupation which corresponds to the mean square displacement in the anisotropic $Pnma$ model.

isotropic model with only two additional distortions for $U_{11/33}$ of Ag and $U_{11/33}$ for the F2 position seems to be the most simple one that correctly describes the local disorder. The results obtained with this model are discussed in the main text (and given in Table I), as it implies a smaller number of parameters and therefore is less affected by some artifacts of the refinement.

Also at the higher temperatures, there is no clear preference for any of these models, which involve different numbers of free parameters and which describe the local disorder in different ways. Due to the larger number of parameters the disorder and $P112_1/a$ models are more easily affected by artifacts of the refinement. However we want to show in the following that the main conclusion about the orbital rearrangement is obtained in all three structure models.

In the disorder model, we add an extra F2' and vary only its $x, y, z$ parameters and occupancies while assuming the same $U_{iso}$ for both sites. These refinements yield very similar results, which further support the local disorder. The reliability parameters with this model are also very close to those obtained with the anisotropic $Pnma$ model, see Table II. Fig. 2 gives the temperature dependence of the Ag-F distances in this model. Already at 250 K one obtains an Ag-F1/F2 octahedron that is elongated as expected. Also the Ag-F1/F2' octahedron is elongated but with an elongation direction rotated by 90° with respect to the Ag-F1/F2 one. This model thus perfectly agrees with disorder deduced from the anisotropic $Pnma$ model, and further corroborates that it stems from disorder in the orbital arrangement. However, the stronger distortion of the minority extra site seems unrealistic. Upon cooling we find the same temperature driven inversion of the octahedron elongations considering both F2 and F2'. The minority orbital scheme is orthogonal to the main one also in the low-temperature phase. This is illustrated by the change of the main displacements between the sites, that are $\Delta x$ at low and $\Delta z$ at high temperatures, respectively, see Fig. 3. The occupation of the extra F2' site increases at 250 K at the expense of that at the F2 site. The product of $\frac{1}{4}\Delta x^2$, respectively $\frac{1}{4}\Delta z^2$, (in absolute units) and the extra occupancy follows the increase of the $U_{11/33}$-F2 parameter in the simple $Pnma$ model.

Also in the structural refinements in spacegroup $P112_1/a$ the change of the orbital arrangement is clearly visible, see Fig. 4. The longest and shorted Ag-F distances are inverted at the transition. One, Ag1, develops an enormous splitting of the in-plane differences by 0.5 Å while that at the other site, Ag2, becomes reduced at low temperature. The apparent disorder of the orbital and bond-distance order seems to be described in this model by the alternating layers with strong and weak differences. Although we cannot exclude this model basing on the refinements, the description with the simpler disorder models seems more likely.

The three approaches to describe the disorder in $KAgF_3$ arrive at the same conclusions. The disorder is induced by an imperfect orbital ordering which strongly resembles the observation of two competing phases in

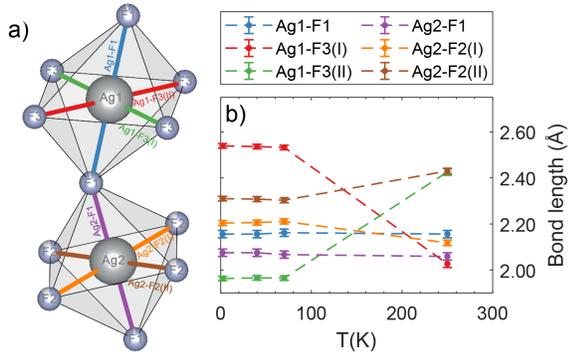

FIG. 4. Sketch of the two distinct AgF$_6$ octahedra existing in the $P112_1/a$ model and temperature dependence of the Ag-F bond distances.

KCuF$_3$, see the discussion in the main text. In all models we find the temperature driven inversion of the orbital ordering across the structural phase transition. Due to the lower amount of parameters the description with the anisotropic $Pnma$ model is the most reliable one, as discussed in the main text.

## B. DFT calculations

All calculations were performed in VASP 5.4.4 suite using projector-augmented-wave method. Core electrons were described using standard VASP pseudopotentials. Strong valence electrons correlation was accounted for within the GGA+U framework using Lichtenstein formalism ($U_{Ag} = 8.0$ eV and $J_{Ag} = 1.0$ eV), and Perdew-Burke-Ernzerhof DFT-GGA functional revised for solids (PBEsol). The cut-off energy of the plane wave basis set was equal to 800 eV, with $k$-point mesh spacing equal to 0.16 Å$^{-1}$. Additional numeric precision was ensured by the use of the Prec=Accurate setting. Geometry relaxations were conducted with default VASP algorithms, and the self-consistent-field convergence criterion was set to $10^{-7}$ eV per unit cell. Both atomic positions and unit cell volume were relaxed in geometric relaxation. After standard geometric relaxation, additional relaxation with regard to forces was conducted reaching forces lower than 0.0005 eV/Å per atom.

### 1. Vibrational properties determined by DFT

Vibrational properties were studied using 2x2x2 supercells derived from well-optimized structures. The frozen phonon method was used, introducing displacements of 0.01 Å by the means of *phonopy* package [1]. Results were also analyzed by *phonopy*, yielding phonon dispersion curves, phonon density of states graphs, irreducible representation assignment to each mode and animation files for each mode allowing for the description of mode char-

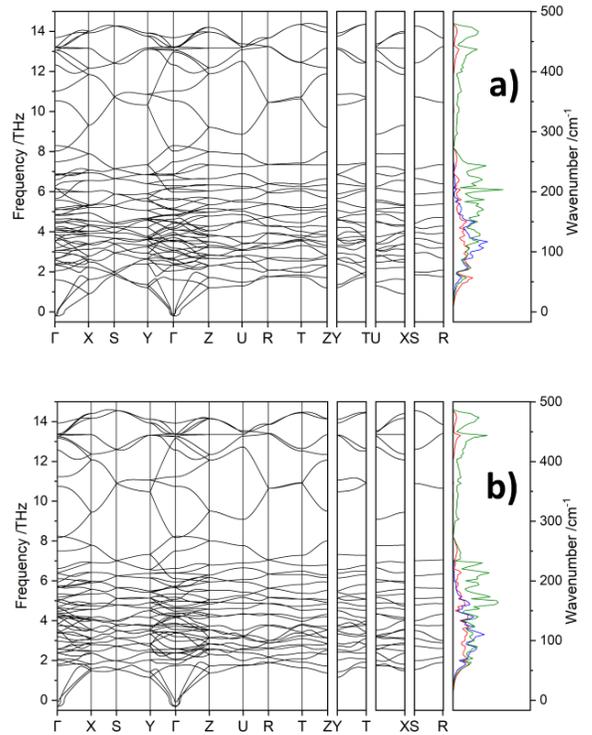

FIG. 5. Dispersion curves (left) and atomically resolved phonon density of states graph (right) for the low-temperature (a) and high-temperature (b) phases of KAgF$_3$ (Ag – red, F – green, K – blue).

acter. The phonon dispersion curves calculated along a standardized $k$ path and the phonon density are shown in Fig. 5.

Analysis of phonon frequencies at the zone center (cf. SI) shows that there are 57 distinct optic phonons for each structure, $\Gamma_{optic}$ = 7Ag + 8Au + 5B1g + 9B1u + 7B2g + 7B2u + 5B3g + 9B3u. Out of those, 25 modes are infrared active (9B1u + 7B2u + 9B3u) while 24 are Raman active (7Ag + 5B1g + 7B2g + 5B3g), so 8 modes are silent (Au). Since both phases are structurally similar, and they adopt the same space group, corresponding phonon frequencies may be compared. It turns out that the twelve highest-frequency Ag-F stretching modes have systematically larger frequencies for the high-temperature phase (by 5 to 8 cm$^{-1}$); this includes modes connected with the stretching of the intra-chain Ag-F bonds. The calculated phonon stiffening is consistent with slight shortening of the intra-chain Ag-F bond lengths in the LT phase as compared to the HT one (i.e., 2.092 vs. 2.095 Å, which is associated with the discussed changes of the crystallographic $b$ vector and the opening of the Ag-F-Ag angle).

Another way of looking at the LT/HT phase transition is related to coordination sphere around K ion. We notice that for the HT phase the coordination sphere of K$^+$ is irregular, with Coordination Number of 7 (average K-F separation is 2.815 Å with the eight very long con-

tact at 3.500 Å). However, the CN for K$^+$ increases to 8 for the LT phase (with the average K-F separation of 2.782 Å). The more compact ligand environment around potassium ion in the LT phase may contribute to the driving force for the phase transition, similarly as for related K$_2$AgF$_4$[2].

### 2. Phase Stability

To understand the driving force of the transition, we analyzed various contributions to the free energy difference between *Pnma* phases. Note that for the related tetragonal CsAgF$_3$ these two minima should be energy degenerate. However, for the orthorhombic system, the degeneracy will be lifted. Indeed, in our DFT calculations, we find that the electronic energy of the most stable magnetic configuration for the LT phase is by 1.8 meV ($\equiv$ 21K) per formula unit lower than for the HT phase. Because the HT phase tends to have slightly softer phonons, the order of the phases inverts if we consider the zero point contribution from lattice vibrations in the harmonic approximation (see above). However, the energy difference is so small ($\lesssim 0.3$ meV) to be indistinguishable from zero within the various methodological errors (e.g. anharmonic effects, DFT deficiencies).

Neglecting the work done against the atmospheric pressure at the structural transition, the Helmholtz free energy should not change, $\Delta F = F_{HT} - F_{LT} = 0$. Since $J$ and the susceptibility both have noticeable changes at the transition, we separate the free energy per FU into a lattice and a magnetic contribution, $\Delta F = \Delta F_{latt} + \delta F_{mag}$. $\Delta F_{latt}$ is estimated computing the temperature dependent free energy of both phases in the harmonic approximation assuming $\Delta F_{latt} = 0$ at $T = 0$. We obtain, $\Delta F_{latt} = -3.4$ meV which is consistent with the fact that entropic effects should favor the HT softer phase.

Regarding the magnetic contribution, since $J_{1D}$ is much larger than other magnetic interactions, the free energy will be dominated by the contributions along the chain. At the transition, $k_B T_s$ is much smaller than $J_{1D}$, so we neglect the spin contribution to the entropy and write the free energy per FU as $F_{mag} = |J_{1D}|\langle \mathbf{S}_0 . \mathbf{S}_b \rangle$ where $\langle \mathbf{S}_0 . \mathbf{S}_b \rangle$ is the correlation between nearest neighbor spins along $b$. Thus, the change of magnetic free energy at the transition is,

$$\begin{aligned} \delta F_{mag} &\approx \delta |J_{1D}|\langle \mathbf{S}_0 . \mathbf{S}_b \rangle + |J_{1D}|\delta \langle \mathbf{S}_0 . \mathbf{S}_b \rangle \\ &\approx 2(\delta |J_{1D}|\langle \mathbf{S}_0 . \mathbf{S}_b \rangle) \\ &\approx 2 \times (-7.4\,\mathrm{meV}) \times (-0.25) \approx 4\,\mathrm{meV} \end{aligned} \quad (1)$$

where, for simplicity, we assumed as an order of magnitude $\delta \langle \mathbf{S}_0 . \mathbf{S}_b \rangle / \langle \mathbf{S}_0 . \mathbf{S}_b \rangle \approx \delta J/J$. The magnetic and the lattice contributions tend to cancel each other. This suggests that the transition is driven by a loss of magnetic energy compensated by a gain on lattice free energy (mainly entropic) of the softer HT phase. In this approximation, the ground state energy of the LT phase should be more stable than the one of the HT phase by 4 meV which is still compatible with the methodological errors of DFT + harmonic corrections. Anharmonic corrections and deficiencies in the DFT functional may explain the ground state energy difference.

 

\end{document}